\documentclass[aps,11pt,pra,showpacs,superscriptaddress,preprint]{revtex4}
\usepackage{amsmath}
\usepackage{graphics}
\usepackage{graphicx}
\usepackage{bm}
\usepackage[pdftex,bookmarks=true,bookmarksopen=true,bookmarksnumbered=true,bookmarksopenlevel=3]{hyperref}
\hypersetup{
    bookmarks=true,     
    unicode=false,          
    pdftoolbar=true,        
    pdfmenubar=true,        
    pdffitwindow=false,     
    pdfstartview={FitH},    
    pdftitle={DRAFT},    
    pdfauthor={Orlando Panella},     
    pdfsubject={Dirac equation with position dependent mass},   
    pdfcreator={latex},   
    pdfproducer={TEXSHOP}, 
    pdfkeywords={Dirac Equation, Position dependent mass }, 
    pdfnewwindow=true,      
    colorlinks=false,       
    linkcolor=red,          
    citecolor=green,        
    filecolor=magenta,      
    urlcolor=cyan           
}

\begin{document}

\title{New exact solution of the one dimensional Dirac Equation
for the Woods-Saxon potential within the effective mass case}

\author{\textsc{O. Panella}}
\affiliation{Istituto Nazionale di Fisica Nucleare, Sezione di Perugia, Via A.~Pascoli, I-06123 Perugia, Italy}

\author{\textsc{S. Biondini}}
\affiliation{Dipartimento di Fisica, Universit\`{a} degli Studi di Perugia, Via A.~Pascoli, I-06123, Perugia, Italy}

\author{\textsc{A. Arda}}
\affiliation{Dipartimento di Fisica, Universit\`{a} degli Studi di Perugia, Via A.~Pascoli, I-06123, Perugia, Italy}
\affiliation{Department of Physics Education, Hacettepe University, 06800, Ankara,Turkey}

\date{\today}

\begin{abstract}
We study  the one-dimensional Dirac  equation in the framework of a position dependent mass under the action of a Woods-Saxon external potential. We find that constraining appropriately the mass function it is possible to obtain a solution of  the problem in terms of the hypergeometric function. The mass function for which this turns out to be possible is continuous.
In particular we study the scattering problem and derive exact expressions for the reflection and transmission coefficients which are compared to those of the constant mass case. For the very same mass function the bound state problem is also solved, providing a transcendental equation for the energy eigenvalues which is solved numerically.
\end{abstract}

\pacs{03.65.-w; 03.65.Ge; 12.39.Fd}

\maketitle


\section{Introduction}

In recent years, the study of several quantum mechanical systems  within the framework of an effective position-dependent mass (PDM) has received increasing attention in the literature. Position dependent effective masses enter, for example, in the  dynamics of electrons in semiconductor hetero-structures~\cite{bastard}, and when describing the properties of hetero-junctions and quantum dots~\cite{PhysRevB.27.7547}.
In non-relativistic quantum mechanics when the mass becomes dependent on the position coordinate the mass  and  momentum operators no longer commute, thereby making  the generalization of the non-relativistic Hamiltonian (kinetic energy operator) to the PDM case highly non trivial~\cite{PhysRev.177.1179,PhysRevB.31.2294}, because of the ambiguities in the  choice of a correct ordering of mass and momentum operator~\cite{PhysRevB.39.12783}. Another important issue is that of Galilean invariance~\cite{PhysRevA.52.1845}.

The investigation of relativistic effects is of course important in those systems containing heavy atoms or heavy ion doping~\cite{Alhaidari:2007nz}. Therefore for these type of materials the investigations of the properties of the Dirac equation in circumstances where the mass becomes a function of the position is certainly of great interest. In addition the problems posed by the ambiguities of the mass and momentum operator ordering are absent in the Dirac equation. An effort in this direction has been reported in some recent literature~\cite{deSouzaDutra2006484,PhysRevA.75.062711,Alhaidari:2007nz,Vakarchuck:2010,Ikhdair:2010dq,Ikhdair:2010dk,Egrifes:2007zz,Jia20091621,Jia2008566}.  For example the authors of ref.~\cite{Alhaidari:2007nz} have reported an interesting \emph{numerical} investigation of the scattering problem for the three-dimensional Dirac hamiltonian within a position dependent mass with a costant asymptotic limit, studying the energy resonance structure. In Ref.~\cite{dekar:2551} the scattering problem  is solved for a smooth potential and a mass step but in the non-relativistic regime. The authors of ref.~\cite{Peng2006478} reported an \emph{approximated} solution of the one-dimensional Dirac equation with a position dependent mass for the generalized Hulth\'{e}n potential. To the best of our knowledge few attempts have been reported that study  the Dirac equation in an external potential with position dependent effective masses. In ref.~\cite{Alhaidari:2004}  the author studies Dirac equation in 3+1 dimension in the Coulomb field and with a spherically symmetric singular mass distribution. In Ref~\cite{Vakarchuck:2010}  the author reports  an exact solution for the Dirac equation with central potential and mass distribution both inversely proportional to the distance from the center.

It is worth pointing out that graphene (single atomic layer of graphite), a recently discovered material~\cite{Novoselov10222004,novoselov-2005-438} which is receiving a lot of attention, exhibits several properties whose explanation involve the Dirac equation for massless fermions. For a comprehensive  review see for example~\cite{RevModPhys.81.109} 
Recent reports studying these effects~\cite{Li2009769,Setare20101433}  attest the use of the Dirac equation in explaining the properties of single-layer graphene .  

The present work is an attempt in the same direction of ref.~\cite{Alhaidari:2004,Vakarchuck:2010}. We report an exact solution of the one-dimensional Dirac equation in the position-dependent mass formalism for a particle  in the Woods-Saxon (WS) potential. Our approach is based on that of Ref.~\cite{Kennedy:2001py} where the author solves the one-dimensional Dirac equation in the Woods Saxon potential for the ordinary \emph{constant} mass case. Our method consist in requiring, within the effective position dependent mass, that the second order equation still be exactly solvable by the hypergeometric function. This is done imposing restrictive conditions on the mass function which lead to a first order differential equation which provides  the explicit mass function. 

We would also like to stress that our new analytical exact solution of the position dependent mass Dirac equation in the Woods-Saxon potential will prove certainly of use in further studies of effective mass models. Other issues could for example be addressed that go beyond the scope of the present work: for example it would certainly be of interest to study in the detail the Klein paradox, as well as the issue of zero momentum resonances that support a bound state at $E=-m$, i.e. super-critical states, in the framework of effective masses.
We also note that the study of the transmission coefficient in the case of  two dimensional Dirac equation for massless fermions has been used already to describe the electrical properties of graphene and in particular the possibility to observe the Klein paradox phenomena~\cite{katsnelson-2006-2} in this material.
In addition the authors of \cite{Setare20101433} discuss the case of massless electrons that cross a square barrier  region where they are instead massive, a situation that can simulate a $n\--p\--n$ junction in a graphene nano-transistor.  
We believe that our exact solution derived for the Woods-Saxon potential with an effective position dependent mass, in the limit $aL\gg 1$ the WS potential barrier reduces to a square barrier,  may prove useful to describe such real physical system. Further investigation in this direction is needed but is beyond the scope of the present work.

The plan of the paper is as follows. In Section II, we summarize the basic equations of the problem. In Section~\ref{scattering_problem}, we solve the effective-mass Dirac equation and provide the mass  function for the problem. We study the scattering problem for the potential barrier and  deduce the transmission and reflection coefficients by studying the asymptotic behavior of the wave function when $x \rightarrow \pm \infty$ and the match at $x=0$. In section~\ref{sec_boundstates} we also address the bound states of the problem by turning the Woods-Saxon potential barrier into a Woods-Saxon potential well. We discuss several numerical examples providing the eigenvalues and wave functions corresponding to particular choices of the parameters. Finally we summarize our results and present our conclusions in Section~\ref{sec_conclusions}.

\section{The BASIC EQUATIONS}\label{sec1}
We recall the free Dirac equation (using natural units, $\hbar=c=1$)
\begin{equation}
\left(i\gamma^{\mu}\frac{\partial}{\partial x^{\mu}}-m\right)\Psi(x)=0\,,
\end{equation}
and $\mu=0,1,2,3$ is a space-time index. Considering instead the case of one space dimension it is possible to choose  the gamma matrices $\gamma^x$ and $\gamma^0$ of dimension two and it is customary to set them respectively to the Pauli matrices
$i\sigma_x$ and $\sigma_z$~[9]. Considering a charge particle minimally coupled to an electromagnetic potential, in the absence of the space component of a vector potential, and setting $V(x)= eA_0(x)$ the one-dimensional Dirac equation for a stationary state $\Psi(x,t)= e^{-iEt}\psi(x)$ becomes:
\begin{equation}
\left[\sigma_{x}\frac{d}{dx}-(E-V(x))\sigma_{z}+m\right]\psi(x)=0\,.
\end{equation}
Decomposing the Dirac spinor $\psi(x)$ into upper ($u_1$) and lower ($u_2$)
components: $\psi= \left( {u_1}\atop {u_2} \right)$, gives the coupled equations:
\begin{subequations}
\begin{align}\label{eq_basics_a}
u'_{1}(x)&=-\left[m+E-V(x)\right]\,u_{2}(x)\,,\\
\label{eq_basics_b}
u'_{2}(x)&=-\left[m-E+V(x)\right]\,u_{1}(x)\,,
\end{align}
\end{subequations}
It turns out to be convenient to define two auxiliary components  $\phi(x)$ and $\chi(x)$ in terms of $u_1(x)$ and $u_2(x)$ as in~[9]:
\begin{subequations}
\begin{align}
\label{eq_phia}
\phi(x)&=u_1(x)+iu_2(x)\,, \\
\label{eq_chia}\chi(x)&=u_1(x)-iu_2(x)\,,
\end{align}
\end{subequations}
Using the above definitions and Eqs.~(\ref{eq_basics_a},\ref{eq_basics_b}) we find the first order coupled equations for the components $\phi(x)$ and $\chi(x)$:
\begin{subequations}
\begin{align}
\label{eq_phi}\phi'(x)&=-i m\chi(x) +i \left[ E-V(x)\right]\phi(x)\,,\\
\label{eq_chi}\chi'(x)&=+i m \phi(x) -i \left[E-V(x)\right]\chi(x)\,,
\end{align}
\end{subequations}
which give the second order equations:
\begin{subequations}
\begin{align}
\phi''(x) + \left[ (E-V(x))^2 -m^2 +i V'(x)\right] \, \phi(x) &=-i m' \, \chi(x)\,,\\
\chi''(x) + \left[ (E-V(x))^2 -m^2 -i V'(x)\right] \, \chi(x) &=+i m' \,\phi(x)\,,
\end{align}
\end{subequations}
where we have taken into account the fact that the mass may depend on the position coordinate and prime denotes the derivative with respect to $x$. Eliminating $\chi(x)$ using Eq.~(\ref{eq_phi}), we obtain for $\phi(x)$ the second order equation:
\begin{equation}
\label{eq_secondorder}
\phi''(x)-\frac{m'}{m}\phi'(x) +\left[ (E-V(x))^2 -m^2 +i V'(x) +i
\,\frac{m'}{m}(E-V(x))\right]\phi(x)=0\,,
\end{equation}
Solving this second order differential equation one can obtain the $\chi(x)$ component via Eq.~(\ref{eq_phi}) and then reconstruct the upper $u_1(x)$ and lower $u_2(x)$ components of the complete spinor solution $\psi(x)$. Eq.~(\ref{eq_secondorder}) reduces to the one studied in \cite{Kennedy:2001py} if $m' =0$, i.e. if the mass reduces to a constant. Thus we see that keeping a position dependence in the mass introduces two new terms: one which multiplies $\phi'(x)$ while the other enters the $\phi(x)$ term. These new terms must be appropriately constrained  in order to be able to solve the equation in terms of the Hypergeometric function.

Let us make a final remark before to discuss the details of the computations. The attentive reader might wonder what would happen if one  were to derive a second order equation for the $\chi(x)$ eliminating instead the $\phi(x)$ component and then computing it via Eq.~\ref{eq_chi}. The second order equation for the $\chi(x)$ component turns out to be:
\begin{equation}
\label{eq_secondorderb}
  \chi''(x)-\frac{m'}{m}\chi'(x)+\bigg[(E-V(x))^2-m^2-iV'(x)-i\frac{m'}{m}(E-V(x))\bigg]\chi(x)=0\,.
  \end{equation}
It is easily checked that Eq.~\ref{eq_secondorderb}  can be obtained from Eq.~\ref{eq_secondorder} using  the map $E\to -E$ and $V(x) \to -V(x)$.
This can be interpreted as the negative energy solution corresponding to the charge conjugate particle (antiparticle). Since $V(x)$ is the temporal component of a four-potential the change $V(x) \to -V(x)$ amounts to reversing the charge of the particle. Indeed with $E\to -E$ and $V(x) \to -V(x)$ we have $\chi \to \phi$ ,
and by using  Eq.~\ref{eq_phi},  $\phi \to -\chi$ and by using the inverse of  
Eqs.~(\ref{eq_phia},\ref{eq_chia}) we have in turn $u_1 \to iu_2$ and $u_2\to iu_1$ which amounts, up to inessential phase factors, to the charge conjugation symmetry of the Dirac equation~\cite{Kennedy:2001py}.

\section{EFFECTIVE-MASS DIRAC SCATTERING PROBLEM}
\label{scattering_problem} 
The form of the Woods-Saxon  potential (illustrated in Fig.~\ref{WSandPDMplot}) is given by (see also ref.~\cite{Kennedy:2001py}):
\begin{equation}
V(x)= W\left[
\frac{\theta(-x) }{e^{-a(x+L)}+1} +
\frac{\theta(x)}{e^{a(x-L)}+1} \right]
\end{equation}
where $W$ is a positive parameter  in the scattering problem (potential barrier) and negative in the bound state problem (potential well); $a$ and $L$ are two real and positive parameters. 

\subsection{\texorpdfstring{Solution in the negative region ($\bm{x<0}$).}{Solution in the negative region (x  <0).}}
Using the variable $y=-e^{-\,a(x+L)}$ and the transformation
$\phi=y^{\mu}(1-y)^{-\lambda}f(y)$ used in ~\cite{Kennedy:2001py}
we will show that it is possible to obtain an exact solution in the
form of an hypergeometric function by imposing appropriate
constraints on the mass function. With the above transformation
and using Eq. (8), Eq.~(\ref{eq_secondorder}) becomes:
\begin{eqnarray}
y(1-y)\frac{d^2f(y)}{dy^2} &+&\left[1+2\mu -y(1+2\mu -2\lambda) -\frac{\dot{m}}{m}
\, y(1-y)\right]\frac{d\, f(y)}{dy} \nonumber \\
&+&\left\{ \lambda(1+2\mu) +\frac{1}{y(1-y)}\left[\mu^2(1-y)^2 +\lambda(1+\lambda)y^2
+\phantom{\frac{1}{2}}\right.\right.\nonumber\\
&+&\left. \left.\frac{1}{a^2}\left[(E^2- m^2)(1-y)^2 +W^2 -2EW (1-y) -iayW \right]\right]
\right\}f(y)\nonumber \\
&-& \frac{\dot{m}}{m} \, y(1-y)\left[\frac{\mu}{y}+\frac{\lambda}{1-y}+\frac{i}{ay}
\left( E-\frac{W}{1-y}\right)\right]f(y) =0\, .
\label{fulleq}
\end{eqnarray}
Here and in the following the dot indicates derivation with respect to the transformed variable ($\dot{m}=dm/dy$). In order  to keep the structure of the hypergeometric differential
equation we impose the following condition on this term:
\begin{equation}
-\frac{\dot{m}}{m} \, y(1-y)= \alpha +\beta y
\end{equation}
which has the following solution ($m_0$ integration constant):
\begin{equation}
\label{massnegative_a}
m(y) = m_0 \, \frac{|y-1|^{\alpha +\beta}}{|y|^\alpha}
\end{equation}
With this choice of mass function Eq.~\ref{fulleq} becomes:
\begin{eqnarray}
y(1-y)\frac{d^2f(y)}{dy^2} &+&\left[1+2\mu -y(1+2\mu -2\lambda) +(\alpha+\beta y) \right]
\frac{d\,f(y)}{dy} \nonumber \\
&+&\left\{ \lambda(1+2\mu) +\frac{1}{y(1-y)}\left[\mu^2(1-y)^2 +\lambda(1+\lambda)y^2
+\phantom{\frac{1}{2}}\right.\right.\nonumber\\
&+& \frac{1}{a^2}\left[(E^2- m^2)(1-y)^2 +W^2 -2EW (1-y) -iayW \right]\nonumber \\
&+& \left.\left.(\alpha+\beta\,y)\left[{\mu}{(1-y)}+{\lambda}{y}+\frac{i}{a}
\left( E(1-y)-{W}\right)\right]\right]\right\}f(y) = 0
\label{fulleqb}
\end{eqnarray}
In order that Eq.~\ref{fulleqb} keeps the structure of the hypergeometric differential equation as in the $m=const$ case we may impose the following conditions on the mass function:
\begin{eqnarray}
\label{massconditionnegative}
\lim_{ y\to -\infty} m(y) &=& m_0 \nonumber \\
 m^2(1-y)^2 &=& m_0^2 y^2
\end{eqnarray}
From these conditions we fix completely the two parameters to $\alpha=-1$
and  $\beta=0$ so that the mass function (in the $y \leq 0$ region ) becomes:
\begin{equation}
\label{massnegative}
m(y) = m_0 \, \frac{y}{y-1}
\end{equation}
The  most general condition that we can impose  on the  term
multiplying to $1/[y(1-y)]$ in order that the equation be that of
the hypergeometric function is that it be equal to a constant
$\gamma$. Therefore we get three equations:
\begin{subequations}
\begin{align}
\mu^2 +\frac{E^2}{a^2}+\frac{W^2}{a^2}-2\frac{EW}{a^2} -\mu-\frac{i}{a}(E-W)&=\,\phantom{+}0 \label{conditionsa}\\
-2\frac{E^2}{a^2} +2\frac{EW}{a^2}-i\frac{W-E}{a}-2\mu^2-(\lambda-\mu)&=\,\phantom{+}\gamma\label{conditionsb}\\
\mu^2+\lambda(1+\lambda) +\frac{E^2-m_0^2}{a^2}&=\,-\gamma
\label{conditionsc}
\end{align}
\end{subequations}
From Eq.~(\ref{conditionsa}), it is possible to solve for $\mu$
while $\lambda $ is found summing  Eq.~(\ref{conditionsb}) and
Eq.~(\ref{conditionsc}).  We obtain finally:
\begin{subequations}
\begin{align}
\lambda &= i\, \sqrt{\frac{W^2-m_0^2}{a^2}}\\
\mu&=-i\, \frac{(E-W)}{a}\\
\gamma&= \nu^2-\mu^2-\lambda(\lambda+1)
\end{align}
\end{subequations}
having defined  $\nu=i k/a$ where $k^2 =E^2-m_0^2$. Our
Eq.~(\ref{fulleqb}) becomes the differential equation of the
hypergeometric function
\begin{equation}
y(1-y)\frac{d^2 f(y)}{dy^2}+[2\mu-(1+2\mu-2\lambda)y]\frac{d\,f(y)}{dy}-(\mu-\lambda-\nu)(\mu-\lambda+\nu) f=0,
\label{HGEQb}
\end{equation}
and the general solution is (with $D_1$ and $D_2$ constants):
\begin{equation}
\label{generalless0}
f(y)=D_1\,
_2F_1(\mu-\nu-\lambda,\mu+\nu-\lambda;2\mu;y)+
D_2\, y^{1-2\mu}\,_2F_1(1-\mu-\nu-\lambda,1-\mu+\nu-\lambda;2-2\mu;y)\,.
\end{equation}

\subsection{\texorpdfstring{Solution in the positive region ($\bm{x>0}$)}{Solution in the positive region (x > 0)}}
Let us study the other region in which  $x>0$, from Eq.~\ref{eq_secondorder} using
the variable $1/z=1+e^{a(x-L)}$ and the transformation
$\phi=z^{-\omega}(1-z)^{-\rho}g(z)$, we obtain:
\begin{eqnarray}
z(1-z)\frac{d^2g(z)}{dz^2} &+&\left[1-2\omega -z(2-2\rho -2\omega)
-\frac{\dot{m}}{m} \, z(1-z)\right]\frac{d\,g(z)}{dz} \nonumber \\
&+&\left\{ 2\omega(1-\rho) -\frac{\dot{m}}{m} \, z(1-z)
\left[-\frac{\omega}{z}+\frac{\rho}{1-z}\right] \phantom{\frac{1}{2}}\right. \nonumber \\
&+&\frac{1}{z(1-z)}\left[\phantom{\frac{1}{2}}\omega(\omega+1)(1-z)^2
 +\rho(1+\rho)z^2-\omega(1-z)+z\rho-(2z^2)\rho \right.\nonumber\\
&+&\left. \left.\frac{1}{a^2}\left[(E-Wz)^2-m(z)^2 -iaWz(1-z)-ia(E-Wz)
\frac{\dot{m}}{m} \, z(1-z)\right]\right]\right\}g(z)=0\nonumber\\
\label{diffeqgt0}
\end{eqnarray}
Following a  line of thought similar to the one outlined in the previous subsection we obtain the  mass function 
\begin{equation}\label{masspositive}m(z)= m_0 (1-z)\,.\end{equation}
The final result for the parameters $\omega, \rho$ and $\delta$ (the coefficient introduced requiring that the term multiplying the factor $1/[z(1-z)]$ be a costant) is found to be:
\begin{subequations}
\begin{align}
\rho&=-i\frac{(E-W)}{a} =\mu \\
\omega&=i\,\sqrt{\frac{E^2-m_0^2}{a^2}}= i\,\frac{k}{a} =\,\nu\\
\delta&=\lambda^2-\mu^2 -\nu^2 -2\nu
\end{align}
\end{subequations}
The differential equation of the function $g(z)$ in
Eq.~(\ref{diffeqgt0}) becomes:
\begin{eqnarray}
z(1-z)\frac{d^2g(z)}{dz^2}+[1-2\omega-z(1-2\rho-2\omega)]\frac{d\,g(z)}{dz}-(-\rho-\omega-\lambda)\,(-\rho-\omega+\lambda) \,\,g(z)=0\,,
\end{eqnarray}
with the general solution ($d_1$ and $d_2$ constants):
\begin{equation}
\label{genxgr0}
g(z)=d_1\,
_2F_1(-\rho-\nu-\lambda,-\rho-\nu+\lambda;1-2\nu;z)+
d_2\, z^{2\nu}\,_2F_1(-\rho+\nu-\lambda,-\rho+\nu+\lambda;1+2\nu;z)\,.
\end{equation}
Let us briefly comment on the change of variables chosen in the negative  region $x\in[-\infty, 0]$:  $y=-e^{-a(x+L)}$ and in the positive region $x\in [0, +\infty]$: $1/z=1+e^{a(x-L)}$. This implies clearly that $y\in [-\infty, -e^{-aL}]$ and $z\in [(1+e^{-aL})^{-1},0]$ (which reduce to  
$y\in [-\infty, 0]$ and $z\in [1,0]$ with the assumption $aL\gg 1$). 
The reader might worry that the choice in the negative region is not appropriate because then the Hypergeometric series $_2F_1(a,b,c;y)$ would not be convergent as its radius of convergence is $|y| <1$. 
However we would like to stress that (in the negative region) we use the explicit form of the Hypergeometric series only in vicinity of $x \to 0^-$ ($|y|\ll 1$) (in the scattering problem) remaining well within the radius of convergence ($|y|<1$).  When obtaining the asymptotic expression at $x\to -\infty$ ($y \to -\infty$) we use the asymptotic expansion of the Hypergeometric series given in Eq~\ref{HGexpansion}.
In the positive region the expansion in $x\to 0^+$ $\left(z\to 1^-\right)$ to avoid problems of convergence we use  the continuation identity of the Hypergeometric  function given in Eq.~\ref{continuation}.
We can say that our exact solution does not rely at all on the Hypergeometric series but rather on the properties of the differential equation satisfied by $_2F_1(a,b,c;y)$. If one needs to use the Hypergeometric function outside the radius of convergence of the series it is  necessary to use adequate analytic continuation identities~\cite{Abramowitz}. We expect in any case that using other transformations like for example $y=e^{ax}$ and $z=e^{-ax}$ would in the end not alter our conclusions.  Similar considerations apply as well to the change of variables chosen in the discussion of the bound state problem (see section IV, subsections A and B).

\subsection{Mass function}
\begin{figure}[t!]
\includegraphics[scale=1.3]{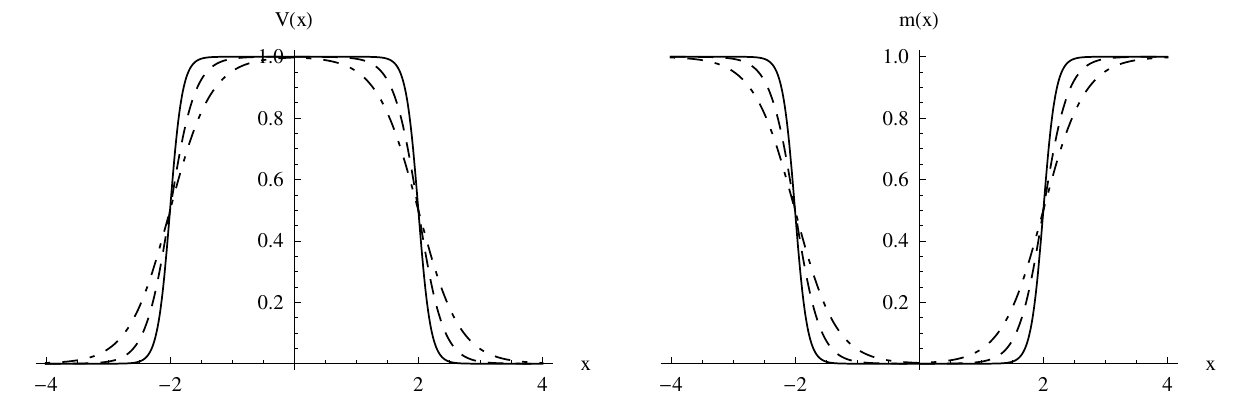}
\caption{\label{WSandPDMplot} Left,  plot of the Woods Saxon potential ($W=1$).
Right, plot of the position dependent mass function $m_0=1$. The parameters are: $L= 2$ and $a=10$ (solid line), $a=5$ (dashed line) and $a=3$ (dot-dashed line). In the limit $aL\gg 1$ the Woods-Saxon  potential reduces to a smooth barrier approaching a square barrier.}
\end{figure}

From conditions given in Eqs.(\ref{massnegative},\ref{masspositive}), we have obtained the mass
function in the two regions as follows in terms of the $x$ variable:
\begin{equation}
\label{massfunction}
m(x)=  m_0 \,\left[ \displaystyle\frac{e^{-a(x+L)}}{e^{-a(x+L)}+1} \, \Theta(-x)
+\displaystyle\frac{e^{a(x-L)}}{e^{a(x-L)}+1}\, \Theta(x)\right]
\end{equation}
Even if the conditions on the parameters $\alpha$ and $\beta$ are
different in the two regions we obtain a continous function,
infact the limit for $x\to 0^{+}$ and $x\to 0^{-}$ is the same;
the value in $x=0$ is $ m(0)=m_0\, {e^{-aL}}/({e^{-aL}+1}) < m_0$
as given in Fig.~\ref{WSandPDMplot}. This mass function has the desired asymptotic
behavior at $x\to \pm \infty$ where it assumes the desired
constant value $m_0$. The reader might be worried that derivatives
of the $\Theta$-functions might introduce singularities in the
problem, since in Eq.~(\ref{eq_secondorder}) there appear terms
with first derivatives of the mass function. However it is quite
straightforward to check that, due to the continuity of the mass
function at $x=0$, such $\delta$-function contributions do cancel
exactly.

An interesting and relevant point  to address comes from the fact that the mass function is of the same type of the vector potential (the Woods-Saxon potential). Indeed the mass function that we have derived (Eq.~\ref{massfunction} can be written in the following form :
\[m(x) = m_0 - \gamma V(x)\] 
where $\gamma= m_0/W$.  Therefore the position dependent mass problem could be looked at as that of a particle of constant mass  $m_0$ coupling to both a vector potential ($V(x)$) and to a scalar potential $S(x)=-\gamma V(x)$. When discussing the barrier ($W>0$) it turns out that $\gamma >0$ and therefore since $S(x)+V(x) =(1-\gamma)V(x)$ there might be regions of the parameter space, $m_0=W$ (or $m_0\approx W$) where   the system is endowed with either an exact (or approximate)  \emph{pseudo-spin symmetry} which is defined by the condition $S(x) +V(x) = constant$. Such symmetry, the near equality of an attractive scalar potential with a repulsive vector potential is well know in the literature~\cite{PhysRevLett.78.436,GinocchioPHYREP} of the Dirac equation and has been proved very useful in  describing the motion of nucleons in the relativistic mean fields resulting  form nucleon-meson interactions, nucleon-nucleon Skyrme-type interactions and QCD sum rules.

Further investigation of the possible consequences of this symmetry for the system under consideration goes however beyond the scope of the present work.

\subsection{Asymptotic expressions and boundary conditions of the scattering problem}
In the negative region from Eq.~(\ref{generalless0}) we have
\begin{eqnarray}
\label{phinegative}
\phi_L(y)&=&D_1\,y^{\mu}\,(1-y)^{-\lambda}
\,_2F_1(\mu-\nu-\lambda,\mu+\nu-\lambda;2\mu;y)\nonumber\\
&+&D_2\,y^{1-\mu}\,(1-y)^{-\lambda}\,_2F_1(1-
\mu-\nu-\lambda,1-\mu+\nu-\lambda;2-2\mu;y)\,.
\end{eqnarray}
We can derive the asymptotic expression as $x\to -\infty$ ($y\to -\infty$) by using the following formula for the asymptotic behaviour of the hypergeometric function~\cite{Abramowitz}
\begin{equation}
\label{HGexpansion}
_2F_1(a,b,c;y)=\frac{\Gamma(c)\Gamma(b-a)}{\Gamma(b)\Gamma(c-a)}\,(-y)^{-a}+
\frac{\Gamma(c)\Gamma(a-b)}{\Gamma(a)\Gamma(c-b)}\,(-y)^{-b}\,,
\end{equation}
and obtain:
\begin{equation}
\phi_L(x)\sim G\, e^{-ik(x+L)}+H\,e^{ik(x+L)}
\end{equation}
where
\begin{subequations}
\begin{align}
G&=D_1 \,A\, e^{i\pi\mu}-D_2\,C\,e^{-i\pi\mu}\\
H&=D_1\, B\, e^{i\pi\mu}-D_2\, D\, e^{-i\pi\mu}
\end{align}
\end{subequations}
and $A,B,C,D$ are given by:
\begin{subequations}
\begin{align}
A&=\frac{\Gamma(2\mu)\Gamma(2\nu)}{\Gamma(\mu+\nu-\lambda)\Gamma(\mu+\nu+\lambda)}\\
B&=\frac{\Gamma(2\mu)\Gamma(-2\nu)}{\Gamma(\mu-\nu-\lambda)\Gamma(\mu-\nu+\lambda)}\\
C&=\frac{\Gamma(2-2\mu)\Gamma(2\nu)}{\Gamma(1-\mu+\nu-\lambda)\Gamma(1-\mu+\nu+\lambda)}\\
D&=\frac{\Gamma(2-2\mu)\Gamma(-2\nu)}{\Gamma(1-\mu-\nu-\lambda)\Gamma(1-\mu-\nu+\lambda)}
\end{align}
\end{subequations}
Similarly we can derive the asymptotic form of lower component  $\chi(x)$ from Eq.~(\ref{eq_phi}):
\begin{equation}
\lim_{x\to -\infty}\chi(x)_{L} = G\,\frac{(E+k)}{m_0}\,e^{-ik(x+L)}+H\,\frac{(E-k)}{m_0}\,e^{ik(x+L)}
\end{equation}
Similarly for the solution in the positive region we have from
Eq.~(\ref{genxgr0}):
\begin{eqnarray}
\label{phipositive}
\phi_{R}(z)&=&d_1\,z^{-\nu}\,(1-z)^{-\rho}
\,_2F_1(-\rho-\nu-\lambda,-\rho-\nu+\lambda;1-2\nu;z)\nonumber\\
&+&d_2\,z^{\nu}\,(1-z)^{-\rho}\,_2F_1(-\rho+\nu-\lambda,-\rho+\nu+\lambda;1+2\nu;z)\,.
\end{eqnarray}
Now we recall that $z \to 0$ when $x\to \infty$ and imposing the
boundary condition of the scattering problem that  in the ($x>0$
region) we only have a wave travelling to the right (only the
transmitted wave) we find:
\begin{equation}
\lim_{x\to+\infty}\phi_{R}(x) = d_1\, e^{ik(x-L)}
\end{equation}
and $\chi_R(x)$ is found again through Eq.~(\ref{eq_phi}) in terms of $\phi_R$:
\begin{equation}
\lim_{x\to+\infty}\chi_{R}(x) = d_1\frac{(E-k)}{m_0}\,e^{ik(x-L)}
\end{equation}

\subsection{\texorpdfstring{Match of solution at $\bm{x=0}$}{Match of solution at x=0}}
So far we have derived asymptotic expressions at $x\to \pm\infty$
for the wave function of the scattering problem in the negative
region, $x<0$, ($\phi_L$),  and in the positive region, $x>0$,
($\phi_R$)  which depend respectively on two, $D_1$ and $D_2$, and
one, $d_1$, unknown constants.  In order to have a physical solution of the scattering problem one needs to match the two solutions $\phi_L$ and $\phi_R$ at $x=0$. This is done by imposing the continuity of the wave function and of its derivative at $x=0$ which gives two conditions and  two of the three unknown
constants can be expressed in terms of the one  left out as the ordinary normalization constant.

We need to find the behavior of  the function $\phi_L(x)$ and $\phi_R(x)$ in the vicinity of $x=0$.

As $x\to 0$ we have $|y|\approx e^{-aL}  \ll 1$ as our only assumption throughout the paper  is that  $aL \gg 1$. Thus $(1-y)^{-\lambda}\approx 1$ and from Eq.~\ref{phinegative} we obtain for $\phi_{L}(y)$:
\[
\phi_L(x) \sim D_1 (-e^{-a(x+L)})^\mu +D_2(-e^{-a(x+L)})^{1-\mu} 
\]
having evaluated the two hypergeometric functions to unity as their argument vanishes, the above equation can be put as:
\begin{equation}
\phi_{L}(x) \sim D_1\,e^{i\pi\mu}\,e^{-a\mu(x+L)}-D_2\,e^{-i\pi\mu}\,e^{-a(x+L)}\,e^{a\mu(x+L)}
\end{equation}
In order to extract the behavior of $\phi_{R}(x)$  as $x\to 0$ we
use the continuation identity of the hypergeometric
function~\cite{Abramowitz}
\begin{eqnarray}
\label{continuation}
\lefteqn{_{2}F_{1}(a,b;c;z)=}\\
&&\frac{\Gamma(c)\Gamma(c-a-b)}{\Gamma(c-a)\Gamma(c-b)}
\,_{2}F_{1}(a,b;a+b-c+1;1-z)\nonumber\\
&&+(1-z)^{c-(a+b)}\frac{\Gamma(c)\Gamma(a+b-c)}{\Gamma(a)\Gamma(b)}
\,\,_{2}F_{1}(c-a,c-b;c-a-b+1;1-z)\,.\nonumber
\end{eqnarray}
Proceeding similarly to the case $x<0$ we find (for $x>0$), that as $x\to 0^+$,  $z\to 1 $, and $1-z\approx e^{a(x-L)} \ll 1$ (we assume $aL\gg1$) and  recalling that $\mu=\rho$, from Eq.~\ref{phipositive} with $d_2=0$ we find:,
\begin{equation}
\label{phiR0}
\phi(x)_{R}\sim d_1\, M\, e^{-a\mu(x-L)}+d_1\, N\, e^{(\mu+1)a(x-L)}
\end{equation}
where:
\begin{subequations}
\begin{align}
M&=\frac{\Gamma(1-2\nu)\Gamma(1+2\mu)}{\Gamma(1+\mu-\nu+\lambda)\Gamma(1+\mu-\nu-\lambda)}\\
N&=\frac{\Gamma(1-2\nu)\Gamma(-1-2\mu)}{\Gamma(-\mu-\nu-\lambda)\Gamma(-\mu-\nu+\lambda)}
\end{align}
\end{subequations}
The match of the two solutions $\phi_L(x)$ and $\phi_R(x)$ is done
by imposing the continuity of the wave function and of its
derivative at $x=0$ which gives:
\begin{eqnarray*}
D_1\,e^{i\pi\mu}\,e^{-\mu aL}-D_2\,e^{-i\pi\mu}\,e^{-(1-\mu)aL}&=&d_1\left[\,M\,e^{\mu aL}+N\,e^{-(\mu+1)aL}\right]\\
-\mu a\, D_1\,e^{i\pi\mu}\,e^{-\mu aL}+(1-\mu)a\,D_2\,e^{-i\pi\mu}\,e^{-(1-\mu)a L}&=&
d_1\left[-\mu a\,M\,e^{\mu aL}+\, N\,(\mu+1)a\, e^{-(\mu+1)a L}\right]\nonumber\\
\end{eqnarray*}
and solving:
\begin{subequations}
\begin{align}
D_1&=\frac{d_1\,e^{-i\pi\mu}}{1-2\mu}\,\left[M\,(1-2\mu)\, e^{2\mu aL} +
2 Ne^{-aL} \right]\\
D_2&=\frac{d_1\,e^{i\pi\mu}}{1-2\mu}\,\left[
N(2\mu+1)e^{-2\mu aL} \right]
\end{align}
\end{subequations} 
We would like to remark  that when solving the Dirac equation the continuity condition at a given boundary ($x=0$ in our case) should be imposed by requiring the match of both the upper and lower spinor components ($u_1(x)$ and $u_2(x)$).  In our second order approach based on the introduction of the auxiliary  components $\phi(x)$ and $\chi(x)$, the derivative of one  of the two ($\phi'$) is connected to  the other ($\chi$) because of Eq.~\ref{eq_phi}. In turns both the initial upper $u_1$ and lower $u_2$ components can be expressed in terms of $\phi$ and $\phi'$. Indeed solving Eq.~\ref{eq_phia} and Eq.~\ref{eq_chia}  and using Eq.~\ref{eq_phi} one finds  for $u_1$ and $u_2$:
\begin{eqnarray*}
u_1(x) &=&\phantom{+}\frac{1}{2}\left[\left(1+\frac{E-V(x)}{m}\right)\phi(x)+\frac{i}{m}\phi'(x)\right]\\
u_2(x)&=& -\frac{i}{2}\left[\left(1-\frac{E-V(x)}{m}\right)\phi(x)-\frac{i}{m}\phi'(x)\right]
\end{eqnarray*}
which shows  how  the matching of the wave function $\phi(x)$ and of its derivative $\phi'(x)$ is totally equivalent to requiring the continuity of $u_1(x)$ and $u_2(x)$. We also remind the reader that this method of matching the solution in $x=0$ has been used also in ref.~\cite{Kennedy:2001py}. The above consideration applies as well to subsection C of section~\ref{sec_boundstates}.

\subsection{Probability current density, reflection and transmission coefficients}
The reader might wonder whether the Dirac equation with a position dependent
mass still has a conserved current. It is well know that a continuity equation
for the current is related to the conservation of probability, or unitarity.  It is quite
straightforward to show that a coordinate dependence of the mass does not bring in any change
in the derivation of the conserved current. This is related to the fact the mass multiplies
the spinor wave function and in deriving the conserved current such terms simply drop out as
in the constant mass case.
The probability current density is given by:
\[
J(x)= \,\bar{\psi}(x) \gamma_x \psi(x) \, = i\, \left[u_1^* (x)u_2(x) -u_2^*(x) u_1(x)\right]
\]
which can be also given in terms of the auxiliary functions $\phi$ e $\chi$:
\[
J(x)= \frac{1}{2} \left[|\phi(x)|^2 -|\chi(x)|^2\right]
\]
With the asymptotic form of the wave function we can compute  the left ($x<0$) $J_L$,
and the right $(x>0$) $J_R$ current density:
\begin{eqnarray}
J_{L}=J_{inc}-J_{refl}&=&|H|^2\,\frac{k(E-k)}{m_0^2}-|G|^2\,\frac{k(E+k)}{m_0^2}\,,\\
J_{R}=J_{trans}&=&|d_1|^2\,\frac{k(E-k)}{m_0^2}\,,
\end{eqnarray}
and we can define the  transmission and reflection coefficients:
\begin{eqnarray}
T&=&\frac{J_{trans}}{J_{inc}}=\frac{|d_1^2|}{|H|^2}\\
R&=&\frac{J_{refl}}{J_{inc}}=\frac{(E+k)}{(E-k)}\,\frac{|G|^2}{|H|^2}
\end{eqnarray}
and from the current conservation $\partial_x J(x) =0 $ it follows
that $\int_{-\infty}^{+\infty}  \partial_x J(x) =
\left[J(x)\right]^{+\infty}_{-\infty} = J_R-J_L =0 $, and
therefore $J_L=J_R$ from which we have the unitarity condition
$R+T=1$. The transmission and reflection coefficients can then be
written as:
\begin{subequations}
\begin{align}
T&=\frac{|1-2\mu|^2}{ \left|\,M\, B (1-2\mu)\, e^{2\mu aL}\, + \,N
\left[B\, (2-2\mu)\, e^{(-2\mu+1)aL}- D\, (2\mu+1)\, e^{-2\mu aL}\right]\right|^{2}}\\
R&=\frac{E+k}{E-k}\frac{|A[M\,(1-2\mu )\,e^{2\mu aL}+N\,(2-2\mu)\,e^{-aL}]
-C[N\,(1+2\mu)\,e^{-2\mu aL}]|^2}{|B[M\,(1-2\mu)\,e^{2\mu aL}+N\,(2-2\mu)\,e^{-aL}]-D[N (1+2\mu)\,e^{-2\mu aL}]|^2}
\end{align}
\end{subequations}

\begin{figure}[t!]
\includegraphics[scale=1.5]{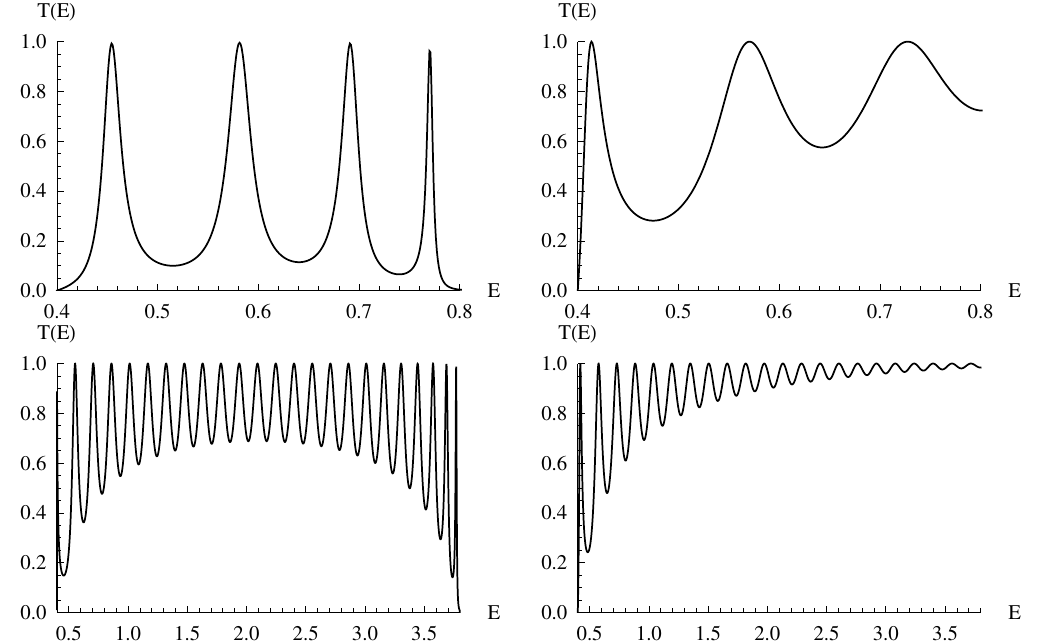}\caption{\label{Tplot} Left,  Transmission coefficient in the
constant mass case. Right: plot of the transmission coefficient in the position
dependent mass case. Parameters: $a=5$, $L= 10$. Upper plots are for  $W=1.2$ and lower plots for $W=4.2$. }
\end{figure}

\begin{figure}[t!]
\includegraphics[scale=1.5]{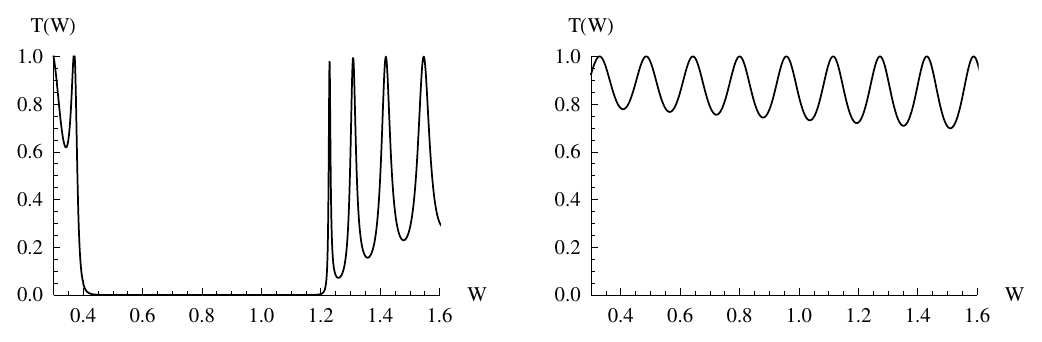}\caption{\label{TWplot}  Transmission coefficient as a function of the barrier height $W$. 
Left: in the constant mass case. Right: the position
dependent mass case. Parameters: $a=5$, $L= 10$, $m_0=0.4$ and $E=2m_0$. }
\end{figure}

{\noindent}Fig.~\ref{Tplot} shows the transmission coefficient in
the constant mass case (left plots) and for the position dependent
mass case (right plot) for two choices of the parameters ($a,L$) and
in the so called Klein range $m <E< W-m$. We note that  in the PDM case
we still observe the transmission resonances found for constant mass. We
observe that, while for $m=m_0$  when $E\to W-m$, $T\to 0$, in the PDM case $T\to 1$. This is
an important  fact  wortwhile to be pointed out. 
In Fig.~\ref{TWplot} we plot the transmission coefficient as a function of the barrier height W. We note that as opposed
to the case of constant mass where $T = 0$ for $E- m < W < E + m$, in the position dependent
mass case, $T(W)$ always oscillates and does not go to zero in the interval $E -m < W < E + m$.
We note also that we have verified our numerical calculations of the constant mass case with those
of ref.~\cite{Kennedy:2001py} finding complete agreement.
Finally, we have numerically
checked the validity of the unitarity condition $R+T=1$.

\section{EFFECTIVE-MASS DIRAC Equation, bound states}\label{sec_boundstates}
Let us study the bound states for the particle with position dependent mass. In order
to do this we take the W-S potential with $W \rightarrow -W$.
\subsection{Negative region}
In the study of the discrete spectrum  it is convenient to use a
different variable. Now we choose $y^{-1}={1+e^{-\,a(x+L)}}$,  with ${d}/{dx}=ay(1-y)\,{d}/{dy}$, $V(y)=-Wy$ 
and $m(y)=m_0\,(1-y)$  and using the parametric transformation
$\phi=y^{\sigma}(1-y)^{\epsilon}h(y)$, we obtain from
Eq.~(\ref{eq_secondorder})
\begin{eqnarray}
&&y(1-y)\frac{d^2h(y)}{dy^2}+[1+2\sigma-y(1+2\sigma+2\epsilon)]\frac{d\,h(y)}{dy^2}+\nonumber\\
&&+\frac{1}{a^2y(1-y)}[(E+Wy)^2 -m_0^2(1-y)^2 -iaWy(1-y) -iay(E+Wy)]\, h(y)\nonumber\\
&&+\frac{1}{y(1-y)}[\sigma(\sigma-1)(1-y)^2+\epsilon(\epsilon-1)y^2+\sigma(1-y)^2]h(y)
+(-2\epsilon\sigma-\epsilon)h(y)=0
\end{eqnarray}
The  most general condition that we can impose  on the  term
multiplying $1/[y(1-y)]$ in order that the equation be that of
the hypergeometric function is that it be equal to a constant
$\zeta$. Therefore we get three equations:
\begin{subequations}
\begin{align}
\sigma^2 +\frac{E^2}{a^2}-\frac{m_0^2}{a^2}&=\,\phantom{+}0 \label{conditionsA}\\
2\frac{m_0^2}{a^2} +2\frac{EW}{a^2}-i\frac{W+E}{a}-2\sigma^2&=\,\phantom{+}\zeta\label{conditionsB}\\
\sigma^2+\epsilon^2-\epsilon+\frac{W^2}{a^2}-\frac{m_0^2}{a^2}&=\,-\zeta
\label{conditionsC}
\end{align}
\end{subequations}
From Eq.~(\ref{conditionsA}) we can solve for $\sigma$, while
summing  Eq.~(\ref{conditionsB}) and  Eq.~(\ref{conditionsC}) we
obtain the equation for $\epsilon$
\begin{subequations}
\begin{align}
\sigma &=\frac{\sqrt{m_0^2-E^2}}{a} \\
\frac{(E+W)^2}{a^2}-i\frac{W+E}{a}-\epsilon+\epsilon^2&=\,0
\end{align}
\end{subequations}
So $\sigma=\nu$. A solution of the second one is
$\epsilon=-i{(E+W)}/{a}$, which is the same of $\mu$ in the
scattering problem but with the replacement $W \rightarrow -W$.  In determining the
parameters $a$ and $b$ ($c=1+2\nu$) of the hypergeomtric equation
we make use  of the relation in  Eq.~(\ref{conditionsC}), where we
define $\lambda=i\, \sqrt{{(W^2-m_0^2)}/{a^2}}$  and finally
obtain:
\begin{equation}
y(1-y)\frac{d^2f(y)}{dy^2}+[1+2\nu-(1+2\epsilon+2\nu)y]\frac{d\, f(y)}{dy^2}-(\epsilon+\nu+\lambda)(\epsilon+\nu-\lambda) f(y)=0.
\label{HGEQBB}
\end{equation}

\subsection{Positive region}

We consider the same substitution of scattering states for the
variable $x$ in the positive region. So we use the variable
$1/z=1+e^{a(x-L)}$, and the transformation
$\phi=z^{-\tau}(1-z)^{-\eta}g(z)$
\begin{eqnarray}
\label{positiveregionbound}&&z(1-z)\frac{d^2g(z)}{dz^2}+[1-2\tau-z(1-2\eta-2\tau)]\frac{d\,g(z)}{dz}+ \nonumber\\
&&+\frac{1}{a^2z(1-z)}[(E+Wz)^2 -m_0^2(1-z)^2 +iaWz(1-z) +iaz(E+Wz)]g(z) \nonumber\\
&&+\frac{1}{z(1-z)}[\tau(\tau+1)(1-z)^2+\eta(\eta+1)z^2-\tau(1-z)+\eta z -2\eta z^2-
\tau z(1-z)+\eta z^2]g(z)\nonumber\\
&&+(2\tau-2\tau\eta)g(z)=0
\end{eqnarray}
Following a line of thought similar to that outlined in the previous subsection ($x<0$) we obtain $\tau^2=-{(E^2 -m_0^2)}/{a^2}=\sigma^2=\nu^2$ and  $\eta=-i{(E+W)}/{a}=\epsilon$ so that Eq.~\ref{positiveregionbound} reduces to:
\begin{equation}
z(1-z)\frac{d^2f(z)}{dz^2}+[1-2\nu-(1-2\epsilon-2\nu)\,z]\frac{d\, f(z)}{dz^2}-(-\epsilon-\nu+\lambda)(-\epsilon-\nu-\lambda) f(z)=0.
\label{HGEQBB1}
\end{equation}

\subsection{\texorpdfstring{Bound state wave function and match at $\bm{x=0}$}{Bound state wave function and match at x=0}}
We note that  the wave function in the $x>0$ region can be obtained from that of the $x<0$ region simply letting $\nu \rightarrow -\nu$ and $\epsilon \rightarrow -\epsilon$. The general solutions to Eqs.~(\ref{HGEQBB},\ref{HGEQBB1}) are:
\begin{eqnarray*}
h(y)&=&A'\,\,
_2F_1(\epsilon+\nu+\lambda,\epsilon+\nu-\lambda;1+2\nu;y) + B'\, y^{-2\nu}\,_2F_1
(\epsilon-\nu+\lambda,\epsilon-\nu-\lambda;1-2\nu;y)\,,\\
g(z)&=&C'\,\,
_2F_1(-\epsilon-\nu+\lambda,-\epsilon-\nu-\lambda;1-2\nu;z)+
D'\, z^{2\nu}\,_2F_1(-\epsilon-\nu+\lambda,-\epsilon-\nu-\lambda;1+2\nu;z)\,.
\end{eqnarray*}
Recall the parametric transformation for $\phi_{L,R}$: $\phi_R=z^{-\nu}(1-z)^{-\epsilon}g(z)$ and $\phi_L=y^{\nu}(1-y)^{\epsilon}h(y)$ and that in the limit of $x
\to \pm \infty$ the variable $y \to 0$ as well as  $z \to 0$. Therefore imposing the boundary condition of a bound state (vanishing wave function at infinity) we obtain $B'=C'=0$ and we are left with:
\begin{eqnarray*}
\phi_{L}(y)&=&A'\,\,y^{\nu}(1-y)^{\epsilon}\,\, _2F_{1}(\epsilon+\nu+\lambda, \epsilon+\nu-\lambda, 1+2\nu;y)\\
\phi_{R}(z)&=& D'\,\,z^{\nu}(1-z)^{-\epsilon}\, \,_2F_{1}(-\epsilon+\nu+\lambda, -\epsilon+\nu-\lambda, 1+2\nu;z)
\end{eqnarray*}
With the help of the continuation formula of the Hypergeometric function \cite{Abramowitz} we can extract the behavior of the solution in the vicinity of $x=0$ (recall that for $ x\to 0$, $y,z\to 1$ and $1-y\approx e^{-a(x-L)}$ while $1-z\approx e^{a(x-L)}$) :
\begin{eqnarray*}
\lefteqn{\phi_L(x)\approx } \\&&A'\, \left\{ \frac{\Gamma(1+2\nu)
\Gamma(1-2\epsilon)}{\Gamma(1-\epsilon+\nu-\lambda)\Gamma(1-\epsilon+\nu+\lambda)} \,
e^{-\epsilon a (x+L)}+\frac{\Gamma(1+2\nu)\Gamma(-1+2\epsilon)}{\Gamma(\epsilon+\nu+\lambda)
\Gamma(\epsilon+\nu-\lambda)}\, e^{-(1-\epsilon) a (x+L)} \right\}\, ,\\
\lefteqn{\phi_R(x)\approx} \\&& D'\, \left[ \frac{\Gamma(1+2\nu)
\Gamma(1+2\epsilon)}{\Gamma(1+\epsilon+\nu+\lambda)\Gamma(1+\epsilon+\nu-\lambda)}\,
e^{-\epsilon a (x-L)}+\frac{\Gamma(1+2\nu)\Gamma(-1-2\epsilon)}{\Gamma(-\epsilon+\nu+\lambda)
\Gamma(-\epsilon+\nu-\lambda)}\, e^{(1+\epsilon) a (x-L)}\right]\,.
\end{eqnarray*}
Upon defining as $S,T,U,V$ respectively the various combinations of gamma functions appearing the above expressions the wave functions are written as:
\begin{subequations}
\begin{align}
\phi_{L}(x)&\approx A'\left[S\, e^{-\epsilon a(x+L)} +T\,e^{-(1-\epsilon)a(x+L)}\right]\\
\phi_{R}(x)&\approx D'\left[U\, e^{-\epsilon a(x-L)}+V\,  e^{(1+\epsilon)a(x-L)}\right]
\end{align}
\end{subequations}
Now we have to match the two solutions in $x=0$ requiring continuity of the wave-function $\phi_L(0) = \phi_R(0)$ and of its first derivative $\phi'_L(0) = \phi'_R(0)$. This gives the homogeneous system:
\begin{eqnarray*}
A'\left[ S\, e^{-\epsilon aL} + T e^{-(1-\epsilon) a L}\right] -D'\left[ U\,
e^{+\epsilon aL} + V e^{-(1+\epsilon) a L}\right] &=&0\\
A'\left[ -\epsilon S\, e^{-\epsilon aL} -(1-\epsilon)  T e^{-(1-\epsilon) a L}\right]
-D'\left[ -\epsilon U\, e^{+\epsilon aL} + (1+\epsilon) V e^{-(1+\epsilon) a L}\right] &=&0
\end{eqnarray*}
which admits a solution only if its determinant is zero. This provides a condition for extracting the energy eigenvalue:
\begin{equation}
\label{eigenvalues}
{\cal F}(E)=\frac{SV}{TU}- e^{4\epsilon aL}\,\, \frac{2\epsilon-1}{2\epsilon+1}\, =0\, .
\end{equation}
When Eq.~\ref{eigenvalues} is satisfied the relation between $A'$ and $D'$ is found to be:
\[
D'= \frac{T}{V}\, \frac{2\epsilon -1}{2\epsilon+1} \, e^{2\epsilon aL} A'\, =\,
\frac{S}{U}e^{- 2 \epsilon a L} \, A' \, ,
\]
and $A'$ is the usual normalization constant.
\begin{figure}[t!]
\includegraphics[scale=1.3]{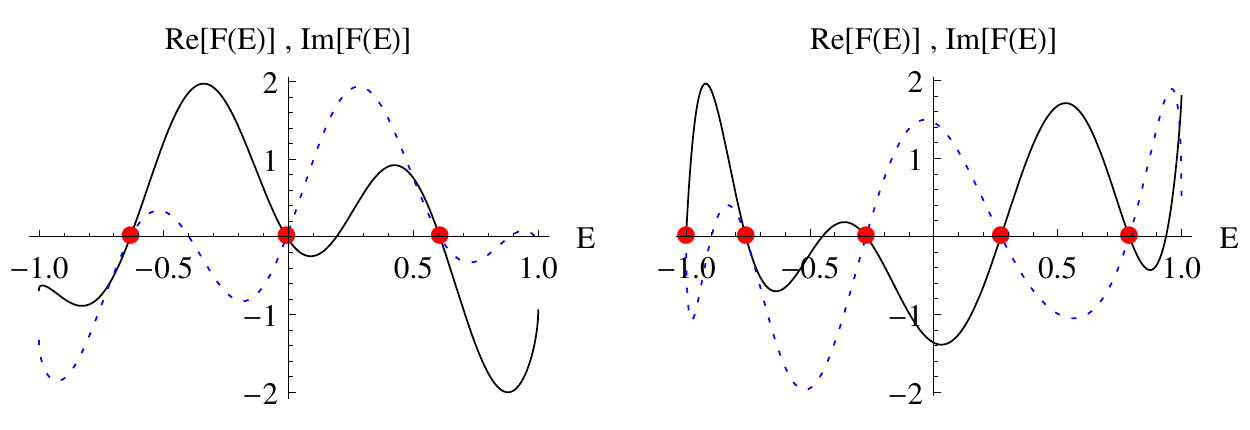}\caption{\label{transcendental}  The energy spectrum is derived by solving numerically with respect to the real variable $E$,  the transcendental equations Re$[{\cal F}(E)]=0$, Im$[{\cal F}(E)]=0$. In the figure we plot Re$[{\cal F}(E)]$ (\emph{solid line}) and Im$[{\cal F}(E)]$ (\emph{dotted line}) as functions of the energy $E$ for the position dependent mass case (left panel) and for the constant mass case (right panel). The eigenvalues (full disks) are given by the points on the $E$-axis where the two curves cross.  The corresponding numerical values  are given in table~\ref{table}.}
\end{figure}
The condition in Eq.~(\ref{eigenvalues}) is a transcendental equation which can be solved numerically.  We provide numerical examples of the bound states. As we are studying bound states we seek numerical solutions of Eq.~(\ref{eigenvalues}) in the interval $-W\leq E\leq m$.  Since ${\cal F}(E)=0$ is complex the energy eigenvalues $E$ are found by solving numerically, for real solutions, the two (independent)  equations Re$[{\cal F}(E)]=0$ and  Im$[{\cal F}(E)]=0$.  Fig.~\ref{transcendental} shows graphically the details of the numerical computations both for the position dependent mass and the constant mass cases. 
\begin{table}[b!]
\begin{tabular}{||c|l|l|l|l||l|}\hline\hline
& $E_1$& $E_2$&$E_3$&$E_4$&$E_5$\\
\hline
$m=const.$&$-1$ & $-0.759003$& $-0.273555$& $+0.271144$&$+0.788942$\\
\hline
$m(x)$&$-0.633251$&$-0.00806737$& $+0.605869$&-&-\\
\hline\hline
\end{tabular}
\caption{\label{table} Numerical values of the energy eigenvalues (discrete spectrum) for the bound states. The model parameters are: $m_0=1$, $W=2$, $a=10$ and $L= 2$. In the upper row we have the spectrum of the constant mass case, while in lower row we have the position dependent mass case.  }
\end{table}
In table~\ref{table} we give the numerical results for the spectra of the position dependent mass case and that of the constant mass case, with the same values of the parameters ($m_0 = 1$;$L=2$; $a = 10$; $W = 2$) in order to make a meaningful comparison between the two cases. For the case of the constant mass we have used the results of~\cite{Kennedy:2001py}. We observe that the in the position dependent mass case the number of bound states decreases relative to the constant mass. 
\begin{figure}[t!]
\includegraphics[scale=1.325]{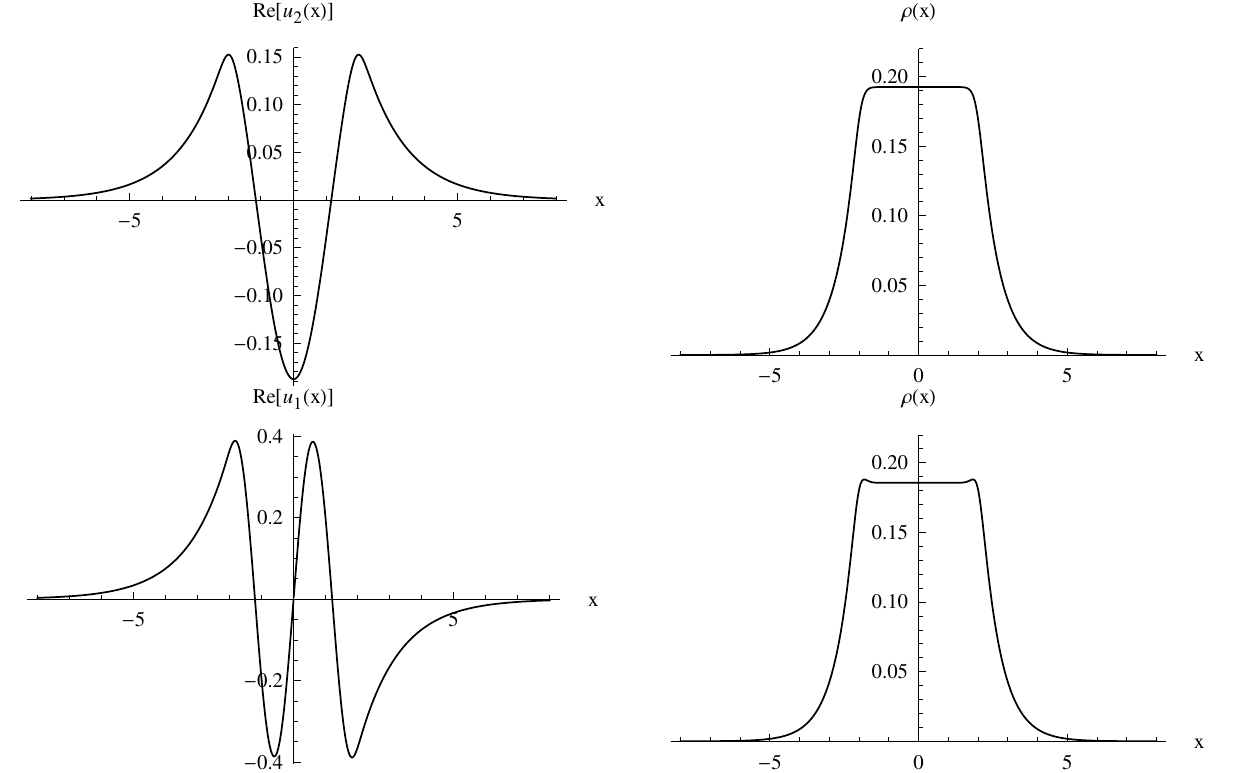}\caption{\label{PDMbound} The case of position dependent mass. Upper panel: plot of the normalized wave function $Re[u_1]$ and of the probability density for $E=-0.633251$ (ground state); Lower panel: the same, but for $E=+0.605869$ (highest excited state).   The model parameters are: $m_0=1$, $W=2$, $a=10$ and $L= 2$. }
\end{figure}
\begin{figure}[t!]
\includegraphics[scale=1.5]{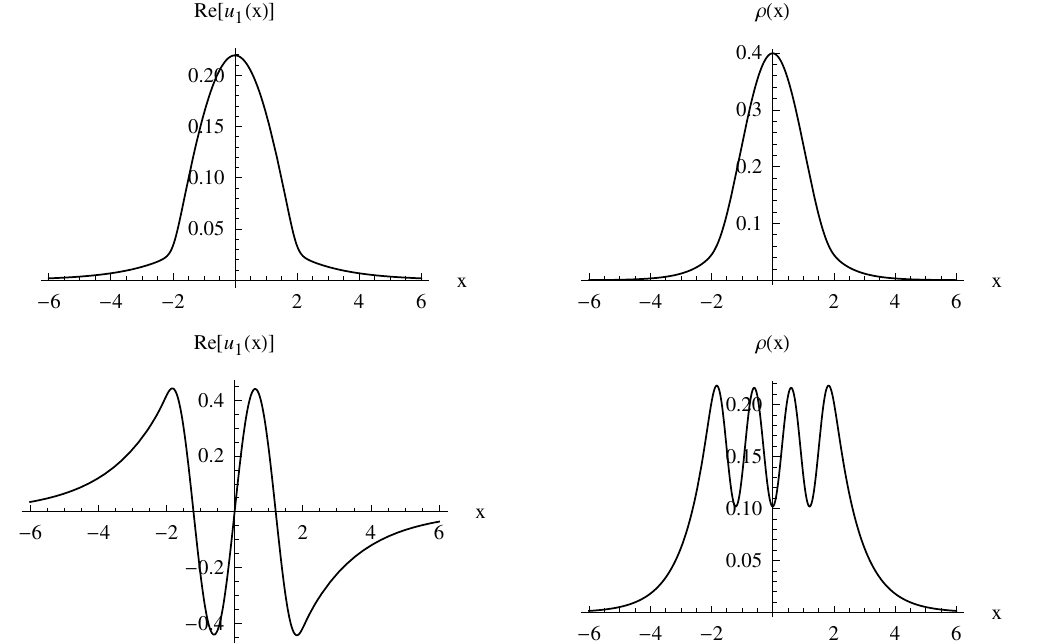}\caption{\label{NONPDMbound} The case of constant mass. Upper panel: plot of the normalized wave function $Re[u_1]$ and probability density for $E=-0.759003$ (lowest energy bound state); Lower panel: the same but for $E=+0.7888942$ (highest excited state).   The model parameters are the same as those of Figure~\ref{PDMbound}: $m_0=1$, $W=2$, $a=10$ and $L= 2$.  }
\end{figure}
In Figures~\ref{PDMbound} and   \ref{NONPDMbound} we provide some example of the (normalized) wave-functions and probability densities both for constant mass case and position dependent mass. Also, comparing Figure~\ref{PDMbound} with Figure~\ref{NONPDMbound} one can infer that in the position dependent mass case the probability density is almost flat in the region inside the potential well as opposed to the constant mass case where, for the highest excited states it oscillates strongly. We note, in the  constant mass case (right panel of Fig.\ref{transcendental}), an eigenvalue corresponding to $E=-m = -1$ which merges with the negative continuum. This situation has previously been considered in the literature~\cite{PhysRevLett.85.1787,PhysRevLett.93.180405,Kennedy:2001py} and has been referred to as super-criticality.  Such super-critical states are also  called half-bound states and are characterized by the fact that one of the spinor components (the upper, $u_1$, or the lower, $u_2$) are not strictly normalizable. We show in table~\ref{table} the eigenvalue $E=-1$ only because it is a solution of  Eq.~\ref{eigenvalues}.  

We do not address further the issue of super-criticality within the position dependent mass case as it goes beyond the scope of the present work.

\begin{figure}[t!]
\includegraphics[scale=1.325]{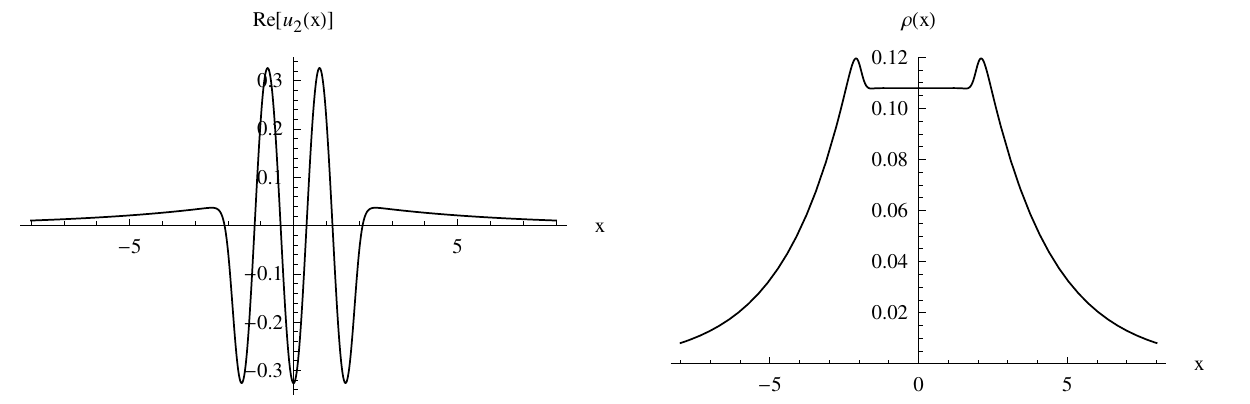}\caption{\label{PDMboundW3} (Position dependent  mass case). Plot of the normalized wave function $Re[u_2]$ (left panel) and of the probability density (right panel) for $E=0.97248$  (highest excited state).   The model parameters are the same as those of Figure~\ref{PDMbound} except for $W$  which now is $W=3$.}
\end{figure}

Finally in Fig~\ref{PDMboundW3} we give a further example in the case of the position dependent mass when the potential well is deeper, $W=3$, with the other parameters as in Figure~\ref{PDMbound}. We note that in this case the highest level is close to the continuum ($E = m_0 = 1$) and indeed the wave function converges less rapidly and the probability density as well. We have computed for example that in this case the probability of the particle to be outside the potential well ($|x| > 2$) is: $P_{outside} \approx 0.57$ which is even greater of the probability to be inside ($-L\le x \le L$) is:  $P_{inside}\approx 0.43$. The less rapid convergence of the wave function is due to the fact that the coefficient that controls such behavior (in this case as $x\to \infty, \phi_R(x) \approx e^{-\nu a x}$) is $\nu= \sqrt{E^2-m_0^2}/a$, and therefore very small giving a wave function that vanishes much slower than those corresponding to the lowest lying bound states.

\section{Discussion and Conclusions}
\label{sec_conclusions} 
We have solved the scattering problem for the one-dimensional Dirac equation with the WS potential in the position-dependent mass formalism. We have set some conditions on the equation in order to keep the structure of the hypergeometric equation which give a suitable mass function. These conditions provide a first order differential equation which can be solved exactly. With the physical requirement that the mass function at infinity goes to a constant mass $m_0$,  specifies completely the mass function. Once the mass function has been found, c.f. Eq.~\ref{massfunction}, we have followed the same technique employed in ref.~\cite{Kennedy:2001py} solving the equation in the negative and positive region separately and giving the solution in terms of the hypergeometric function $_2F_1$. For the scattering problem ordinary boundary conditions at infinity are imposed and then the match at $x=0$ allows to specify all unknowns up to a normalization constant.  

We note that our method of solving  the Dirac equation for a particular case of effective position dependent mass function has the drawback of providing a mass function which does not interpolate smoothly with the constant  mass case. In other words our mass function does not contain a parameter such that  when set to zero reduces the mass function to the constant case ($m_0$).   Further studies in this direction should be pursued in order to overcome such difficulties.

We have obtained analytical expressions for the transmission and reflection coefficients, and we explicitly verified that unitarity ($R+T=1$) is preserved  in the PDM case. We have also studied the bound states, i.e. the discrete spectrum of the WS potential well with the effective position dependent mass, finding an exact analytical condition for the energy eigenvalues (in the form of a transcendental equation which needs to solved numerically). We have provided an explicit numerical example finding the eigenvalues and the wave function for a specific choice of the parameters.

Our approach offers one of the few examples where the Dirac equation is solved exactly in the position dependent mass case and in an external potential.  To the best of our knowledge the only other  example is reported in ref.~\cite{Alhaidari:2004} where the three-dimensional Dirac equation is solved for the PDM case in the Coulomb external field. We note a similarity between the present work and that reported in \cite{Alhaidari:2004}. In both cases the mass function for which an exact solution is found shares similarities with the external potential. In \cite{Alhaidari:2004} the spherically symmetric mass function for which the problem is solved is $m(r) =1 +\mu \lambda^2/r$ where, in atomic units, $m_0=\hbar=1$ and $\lambda$ is the Compton wavelength. Our mass function, c.f. Eq.~\ref{massfunction}, is also certainly related to the shape of the Woods-Saxon  external potential.

\begin{acknowledgments}
This work has grown out of the diploma thesis of S.B. presented at the University of Perugia in September 2009. A.~A. acknowledges warm hospitality from the Physics Department of the University of Perugia and INFN -- Istituto Nazionale di Fisica Nucleare -- Sezione di Perugia.
\end{acknowledgments}

\bibliography{REVISED}{}
\end{document}